\documentclass[twocolumn,prl,superscriptaddress,showpacs,TimesNewRoman]{revtex4}
\usepackage{epsf,graphicx,amssymb,subfig,amsmath}
\usepackage{pslatex}
\begin{document}
\title{Bistability between equatorial and axial dipoles during
  magnetic field reversals}

\author{Christophe Gissinger}
\affiliation{Department of Astrophysical Sciences/Princeton Plasma Physics Lab,
 Princeton University, Princeton NJ USA.  19104-2688}
\author{Ludovic Petitdemange}
\author{Martin Schrinner}
\author{Emmanuel Dormy}
\affiliation{MAG(CNRS/ENS/IPGP), LRA, Ecole Normale Sup\'erieure, Paris Cedex 05, France}

\def\bfnabla{\mbox{\boldmath $\nabla$}}

\begin{abstract}
Numerical simulations of the geodynamo in presence of an heterogeneous
heating are presented. We study the dynamics and the structure of the
magnetic field when the equatorial symmetry of the flow is broken. If
the symmetry breaking is sufficiently strong, the $m=0$ axial dipolar
field is replaced by an hemispherical magnetic field, dominated by an
oscillating $m=1$ magnetic field. Moreover, for moderate symmetry
breaking, a bistability between the axial and the equatorial dipole is
observed. In this bistable regime, the axial magnetic field exhibits
chaotic switches of its polarity, involving the equatorial dipole
during the transition period. This new scenario for magnetic field
reversals is discussed within the framework of the Earth's dynamo.
\end{abstract}
\maketitle

 It is now commonly believed that magnetic fields of the planets,
 including the Earth, are generated by dynamo action due to the fluid
 motion of liquid iron inside their cores \cite{Moffatt}. In most of
 the planets, the magnetic field at the surface is dominated by a
 dipolar magnetic field. In some cases, like the Earth, the dipole
 field is almost aligned with the axis of rotation. But recent
 observations have shown that for some planets, like Uranus or
 Neptune, the dipole axis can be tilted up to $45^o$ due to a
 significant contribution from the equatorial dipole \cite{Jones11}.\\

 In the case of the Earth, paleomagnetic measurements also allow to
 reconstruct the dynamics of the magnetic field. The Earth's dipolar
 field has reversed its polarity several hundred times during the past
 $160$ millions years, and polarity reversals are known to be strongly
 irregular and chaotic. Chaotic reversals have also been reported in
 numerical simulations \cite{Roberts00}, and in a laboratory
 experiment. In the VKS (Von Karman Sodium) experiment, the dynamo
 magnetic field is created by a turbulent von Karman swirling flow of
 liquid sodium due to two counter-rotating bladed disks
 \cite{Monchaux07}. In this experiment, reversals of the axial dipolar
 magnetic field have been reported, but only if the two impellers
 rotate at different frequencies, when the equatorial symmetry of the
 flow is broken \cite{Berhanu07}. These experimental observations are
 in a very good agreement with a recent theoretical model, in which
 reversals arise from the interaction between symmetric and
 antisymmetric components of the magnetic field, linearly coupled by
 the action of an antisymmetric velocity field \cite{Petrelis09},
 \cite{PFD}.

A growing number of studies seem to assess the effect of an
equatorially antisymmetric velocity mode on geomagnetic field
reversals. First, it has been observed that the ends of superchrons
(large periods of time without geomagnetic reversals) are related to
major flood basalt eruptions due to large thermal plumes ascending
through the mantle \cite{Courtillot07}. In agreement with this
observation, it has been shown in geodynamo numerical simulations that
the dipole field reversals and the loss of equatorial symmetry seem to
be tightly connected \cite{Kusano08}, and that taking an heterogeneous
heat flux at the core-mantle boundary of the Earth strongly influences
the frequency of magnetic field reversals \cite{Olson10}. Finally, a
study recently suggested that an equatorially asymmetrical
distribution of the continent is correlated with long term increase of
geomagnetic reversal frequency \cite{Petrelis11}. \\

In this letter, we report 3D numerical simulations of an electrically
conducting, thermally convecting Boussinesq fluid. The fluid is
contained in a spherical shell that rotates about the $z-$axis at the
rotation rate $\Omega$. The boundaries corresponds to fixed
temperature boundary conditions. On the inner sphere of radius $r_i$,
the temperature is homogeneously fixed to $T_i$, but an heterogeneous
temperature pattern $g_1^0$ is used at the outer boundary (of radius
$r_o$). The pattern corresponds to the simplest large scale mode
breaking the equatorial symmetry of the flow:
\begin{equation}
T_o=T_i-\Delta T(1-C\cos\theta)
\end{equation}
where $T_o$ is the temperature at the outer boundary, and $C$ is a
free parameter measuring the amplitude of the equatorial symmetry
breaking. The dimensionless equations system includes the Navier-Stokes equation
coupled to the induction equation and the heat equation, and the
conditions that both magnetic and velocity fields are divergence free.
The dimensionless parameters are the magnetic Prandtl number
$Pm=\nu/\eta$, the Ekman number $Ek=\nu/(\Omega D^2)$, the Prandtl
number $Pr=\nu/\kappa$ and the Rayleigh number $Ra=\alpha g_0\Delta T
D/(\nu\Omega)$, where $D=(r_o-r_i)$ is the typical lenghtscale. $\nu$,
$\eta$, $\kappa$, $\alpha$ and $g_0$ are respectively the kinematic
viscosity, the magnetic diffusivity, the thermal diffusivity, the
thermal expansion coefficient and the gravity at the outer
sphere. Time is expressed in viscous units. The radius ratio is fixed
to $r_i/r_o=0.3$. The inner and outer spheres are electrical
insulators, and no-slip boundary conditions are used on these
boundaries. In all the results reported here, $Ra=120$, $Pm=20$,
$Pr=1$ and $Ek=6.e-3$. Although these parameters are far from those of
natural dynamos, they allow for long time integrations and statistical
analysis \cite{Olson10}. $C$ is varied between $0$ and $0.25$. \\
\begin{figure*}
\centering 
\subfloat[]{
\epsfysize=40mm \epsffile{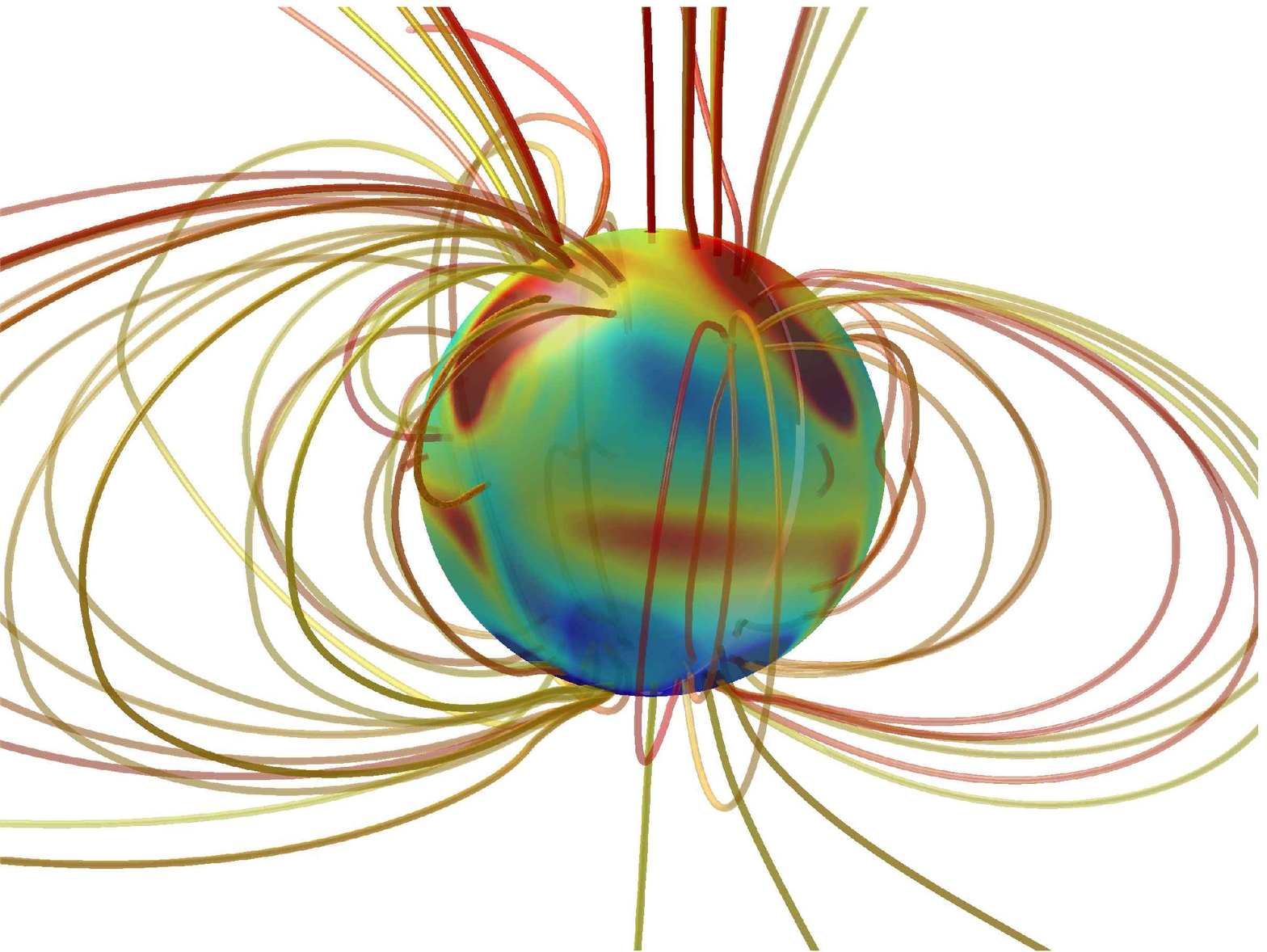}}
\hskip -5mm
\subfloat[]{
\epsfysize=40mm \epsffile{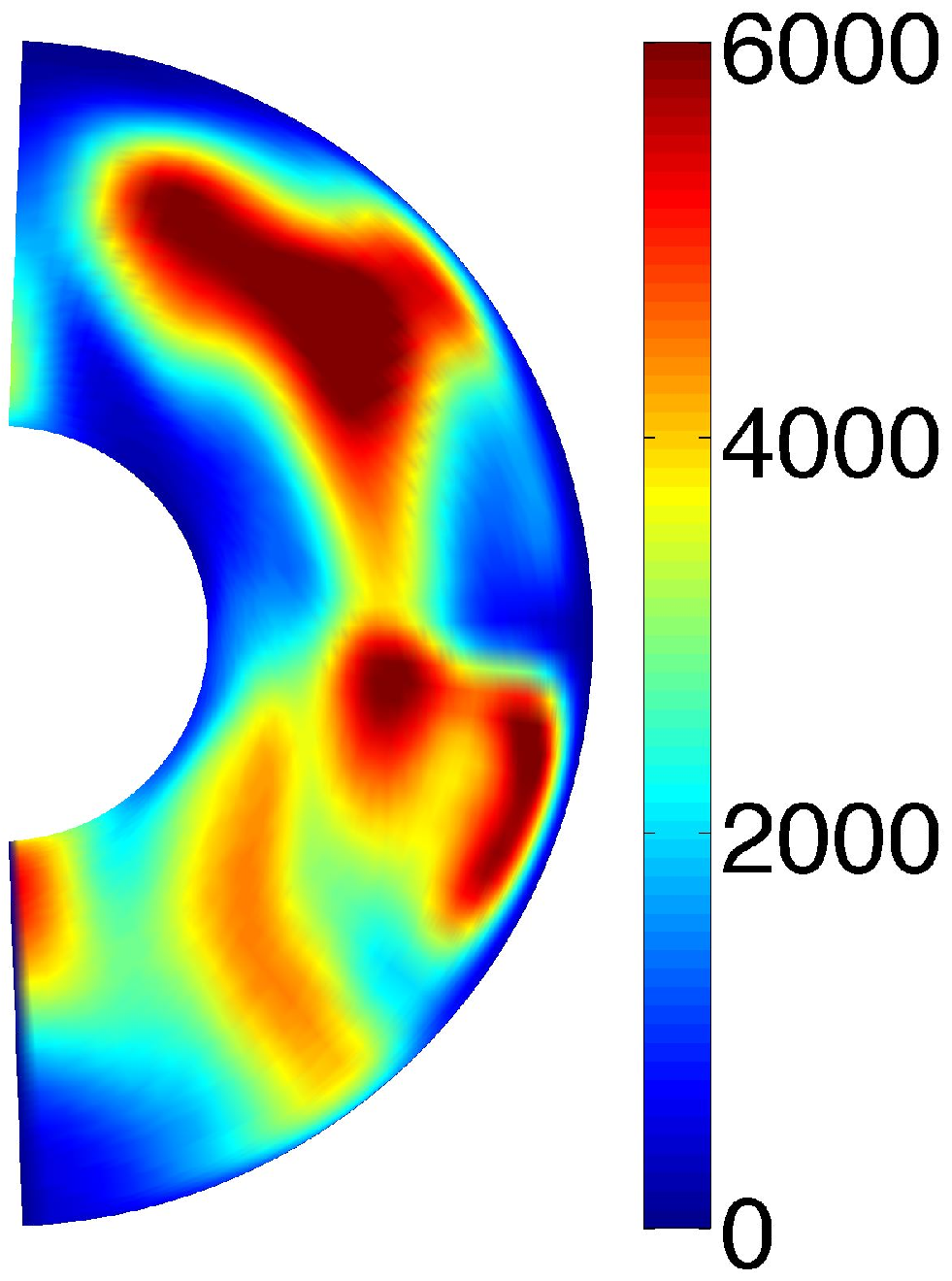}}
\hskip +12 mm
\subfloat[]{
\epsfysize=40mm \epsffile{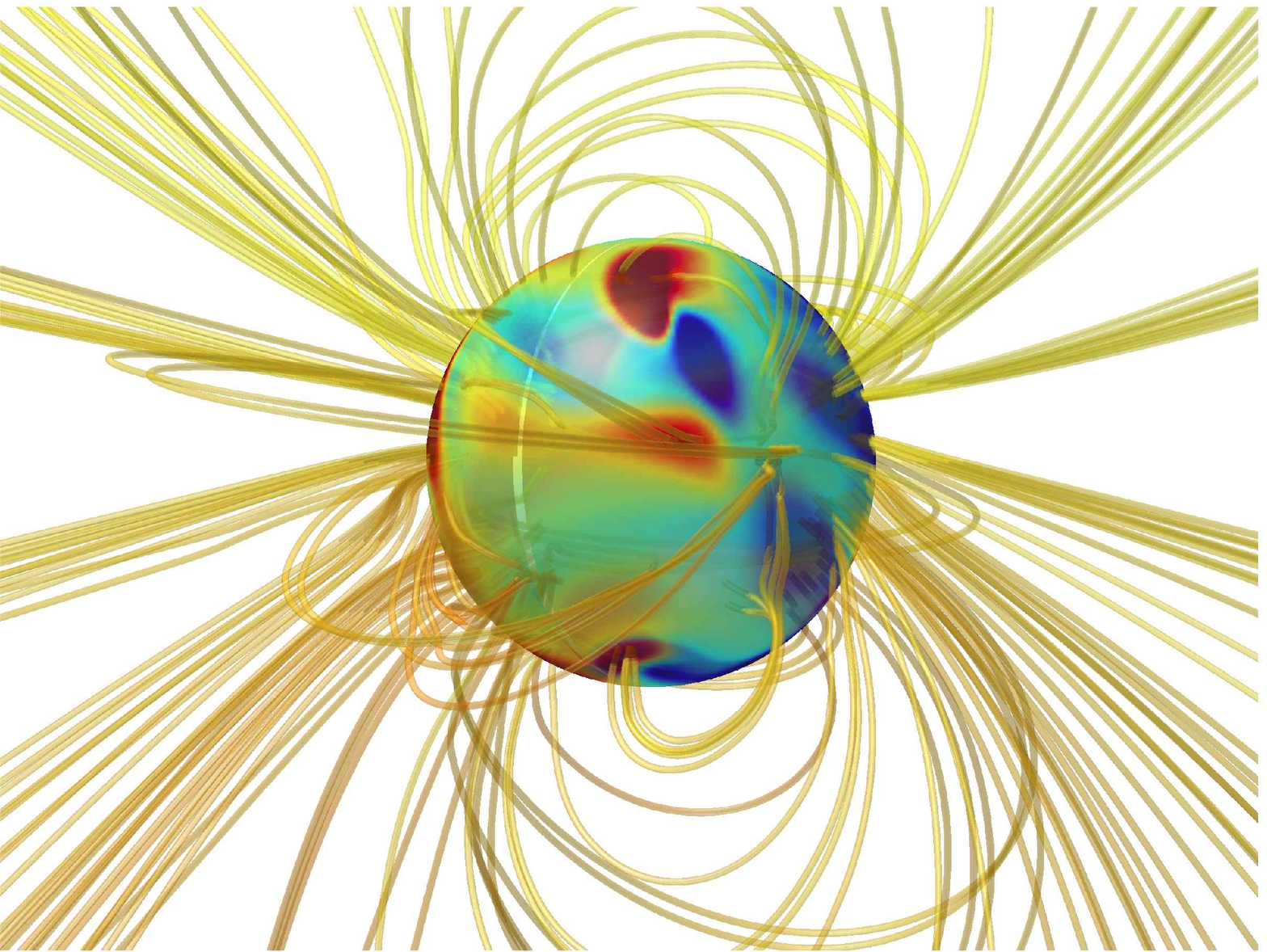} }
\hskip -5mm
\subfloat[]{
\epsfysize=40mm \epsffile{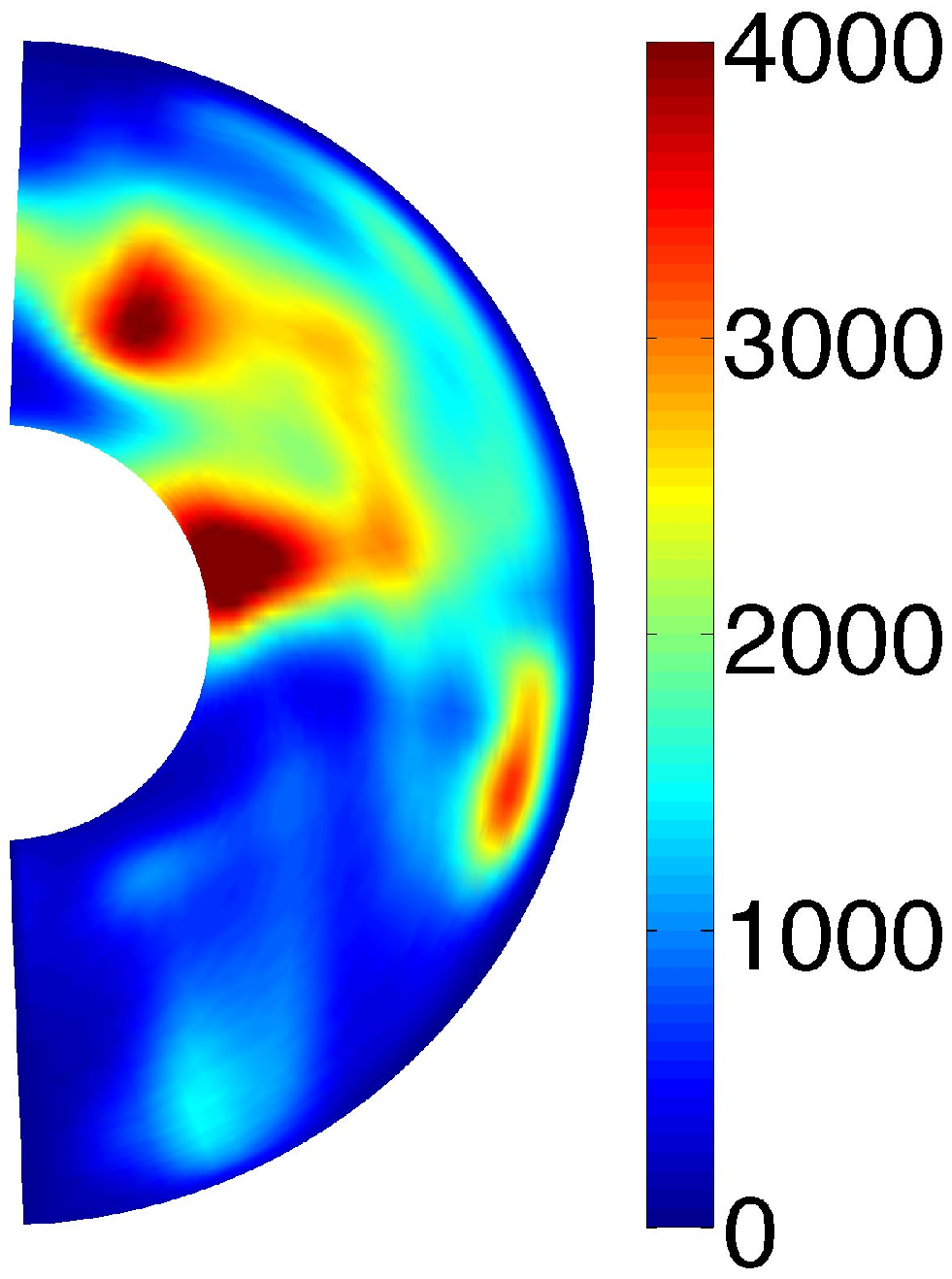}}
\vskip -2mm
\caption{\footnotesize{(a): radial magnetic field $B_r$ at the
    core-mantle boundary for the axial dipolar solution $D$. Magnetic
    field lines are also shown. (b): total magnetic energy
    (azimuthally averaged) in the meridional plane.  The solution $D$
    corresponds to a strong dipolar magnetic field, with a
    distribution of the magnetic energy relatively symmetrical,
    although slightly larger in the northern hemisphere. (b) and (c):
    same thing, but showing the equatorial dipolar solution $E$
    observed at larger values of $C$. The solution $E$ takes the form
    of an equatorial dipole at the outer surface, but the total
    magnetic field in the bulk of the flow is strongly
    hemispherical.}}
\label{solutionDE}
\end{figure*}

 Fig. \ref{solutionDE}a and \ref{solutionDE}b show the solution
 obtained for $C=0.1$, when the symmetry breaking is relatively
 weak. For this value, the magnetic field is strongly dominated by its
 axisymmetric component and the radial magnetic field measured at the
 core-mantle boundary shows a strong dipolar component
 (Fig. \ref{solutionDE}a). A weaker non-axisymmetric component,
 reminiscent from the $m=3$ convection pattern, is also visible. This
 magnetic structure is quite similar to the one obtained in the
 absence of symmetry breaking. Fig. \ref{solutionDE}b represents the
 total magnetic energy averaged in the $\phi$-direction, shown in the
 poloidal plane ($r,\theta$). Despite the heterogeneous temperature
 gradient, the magnetic energy remains largely symmetrical with
 respect to the equator.\\

For larger symmetry breaking, this dipole is replaced by a totally
different solution, hereafter referred as solution
$E$. Fig. \ref{solutionDE}c and \ref{solutionDE}d show the magnetic
structure obtained for $C=0.2$. The magnetic field is now dominated by
a non-axisymmetric $m=1$ component. At the outer sphere, the field
corresponds to an equatorial dipole, rotating around the $z$-axis and
slightly stronger in the northern hemisphere. In the bulk of the flow,
the equatorial asymmetry of the field becomes more important, as shown
by the magnetic energy distribution (Fig. \ref{solutionDE}d), and this
new solution therefore takes the form of an hemispherical magnetic
field (a similar behavior was reported in \cite{Stanley08}). Although
the thermal convection is made more vigorous in the southern
hemisphere by the heterogeneous heating, note that the magnetic energy
is surprisingly localized in the northern hemisphere.

The generation of an equatorial dipole has been reported in previous
numerical studies. An equatorial dipole solution was described for
Rayleigh number very close to the onset of convection
\cite{Ishihara02}, and a similar solution was found in \cite{Aubert04}
for smaller shell thickness. In our case, the breaking of the
equatorial symmetry is directly responsible for the generation of the
equatorial dipole. For the range of $C$ studied here, the total
kinetic energy remains relatively symmetrical with respect to the
equatorial plane (for $C=0.1$, the equatorially antisymmetric flow
energy is only $10\%$ of the symmetrical one). However, this weak
symmetry breaking is sufficient to strongly modify the axisymmetric
velocity, by generating a large counter-rotating zonal flow. This
toroidal $t_2^0$ flow introduces a strong shear in the equatorial
plane which tends to favor the equatorial dipole at the expense of the
axial one.\\
\begin{figure}
\centerline{
\epsfysize=52mm 
\epsffile{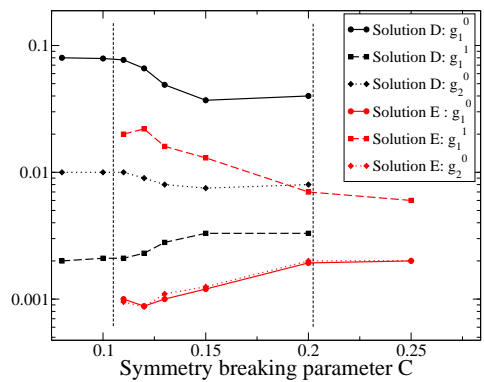} }
\vskip -3mm
\caption{\footnotesize{Bifurcation of the coefficients $g_1^0$,
    $g_1^1$ and $g_2^0$ of the magnetic energy as a function of the
    symmetry breaking parameter $C$. For $0.1<C<0.2$, there is a
    bistability between the axial dipole solution $D$ (black) and the
    equatorial dipole solution $E$ (red). Turbulent fluctuations
    connect the two solutions.}}
\label{bif_bista}
\end{figure}

An interesting behavior occurs for intermediate values of the symmetry
breaking. When $0.1<C<0.2$, a bistability between the axial dipole $D$
and the non-axisymmetric solution $E$ is indeed
obtained. Fig. \ref{bif_bista} illustrates this bistable regime by
showing the bifurcation of both modes as a function of $C$. The axial
dipolar solution $D$ is shown in black, and the solution $E$ dominated
by $m=1$ magnetic modes in red. For each of these solutions, we show
the coefficients of the axial dipole $g_1^0$, the equatorial dipole
$g_1^1$, and the axial quadrupole $g_2^0$, where $g_l^m$ means the
poloidal component of the spherical harmonic of order $l$ and degree
$m$. The dashed vertical lines in Fig. \ref{bif_bista} indicate the
region for which the system is bistable: both solutions can be
obtained depending on the initial conditions of the simulation.  Note
that for the solution $E$, dipolar and quadrupolar components possess
the same amplitude, in agreement with the hemispherical structure of
the magnetic field.\\

\begin{figure}
\centerline{
\epsfysize=45mm 
\epsffile{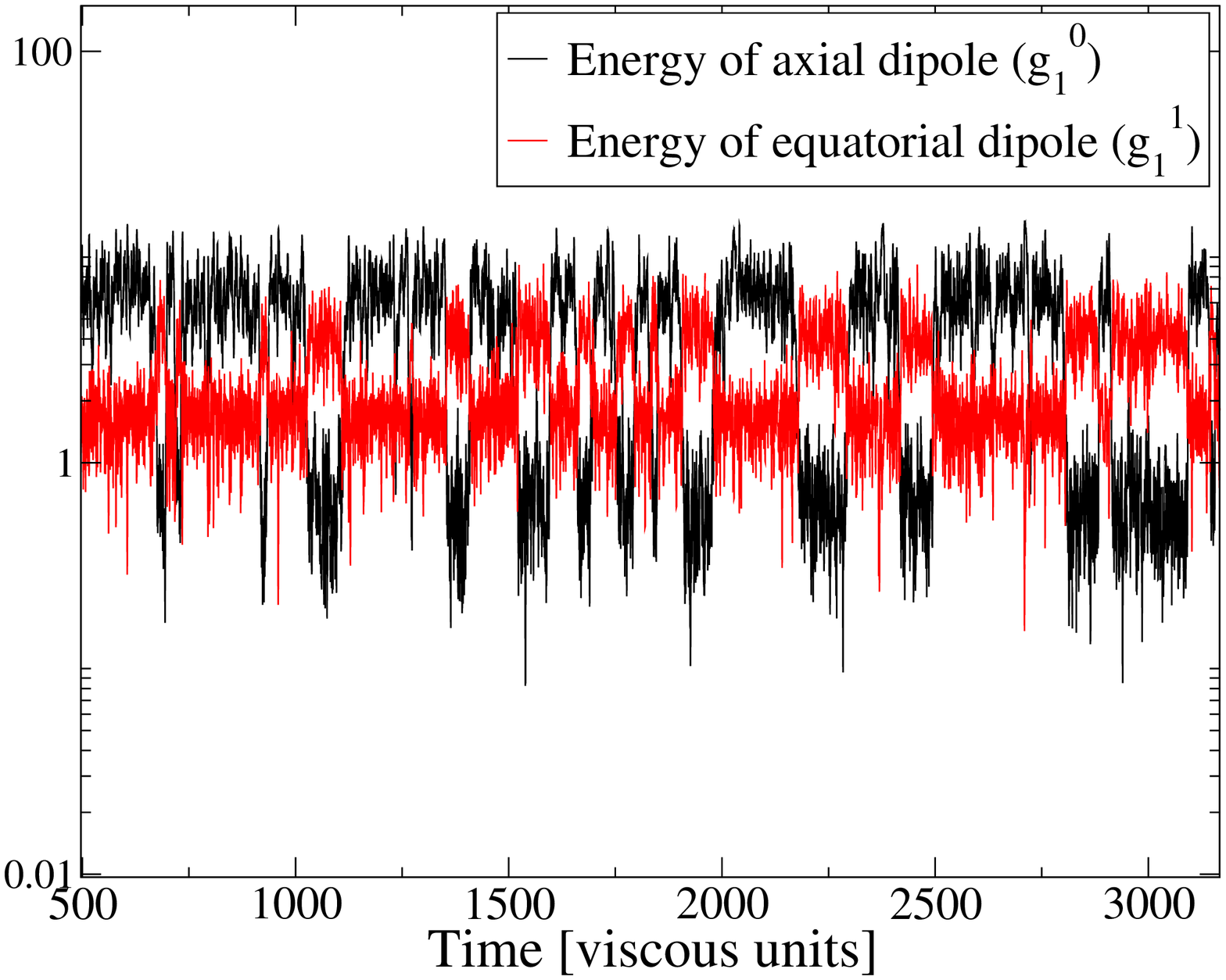}
\epsfysize=45mm 
\epsffile{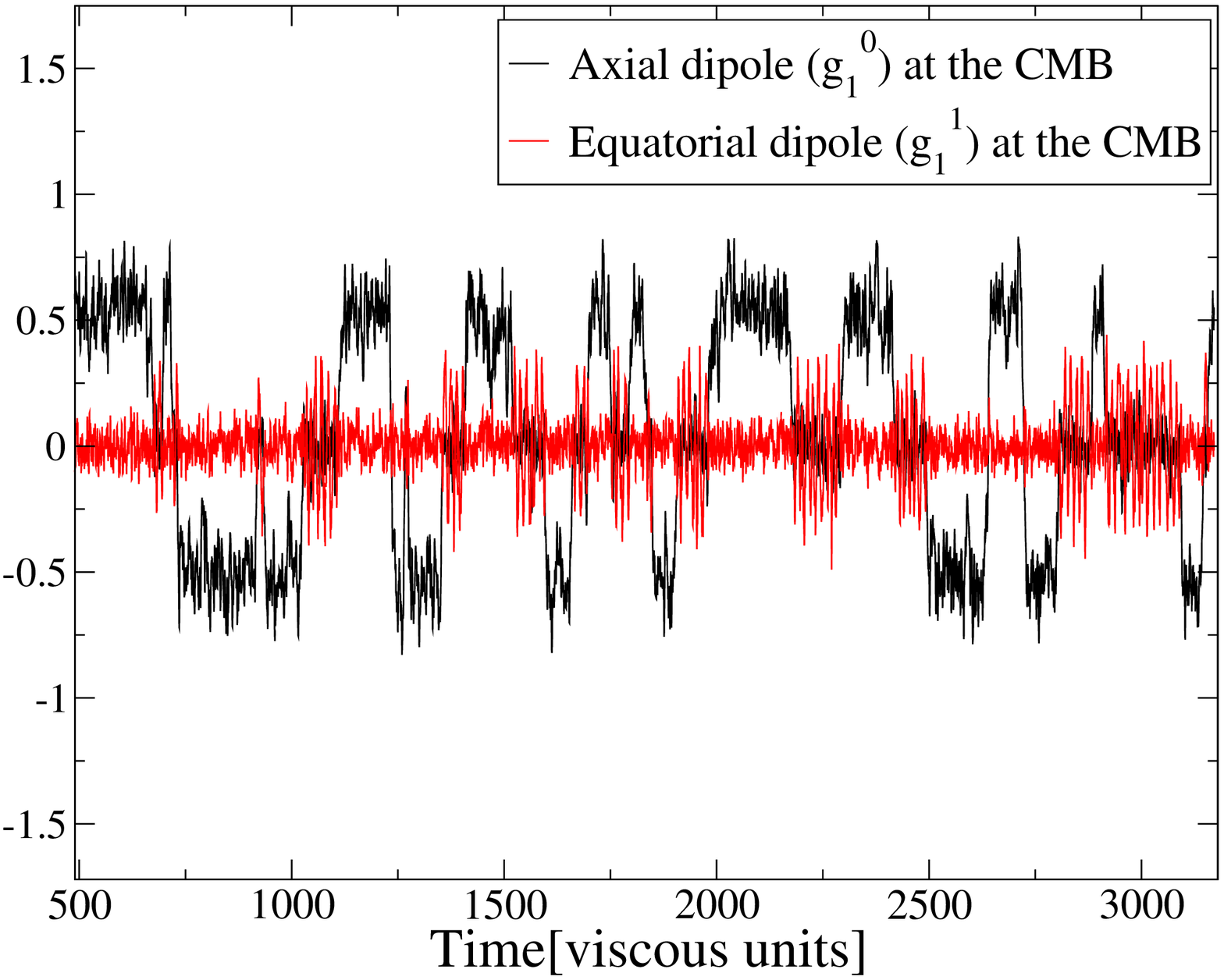} }
\vskip -1mm
\caption{\footnotesize{Time evolution of the magnetic field for
    $C=0.13$. The system chaotically jumps between the bistable
    solutions $E$ and $\pm D$.  Left: Magnetic energy of the
    equatorial (red) and axial (black) dipoles. Right : Same thing,
    but at the core-mantle boundary. The bistability with the
    equatorial dipole yields chaotic polarity reversals of the axial
    dipolar magnetic field.}}
\label{time}
\end{figure}

More interestingly, when the magnetic field is in this bistable
regime, for $0.1<C<0.2$, the strong fluctuations generated by the
turbulence of the flow allow the system to switch from one solution to
the other. These transitions between the axial and the equatorial
dipole are shown by the time series of the energy of the system in
Fig. \ref{time}-left: the two states, although strongly
fluctuating, are clearly distinguishable by different well defined
mean values for the energies of axial (black) and equatorial (red)
dipoles, and the system randomly switch from one state to the
other. In addition, Fig. \ref{time}-right shows the time
evolution of the $g_1^0$ and the $g_1^1$ at the core-mantle
boundary. Since the phase space is symmetrical with respect to the
symmetry $D\rightarrow -D$, we observe transitions from $E$ to $D$ as
well as transitions from $E$ to $-D$. This bistability between
  the axial dipole and the equatorial one therefore takes the form of
  chaotic reversals of the polarity of the axial dipole. During a
reversal, the dipolar magnetic field does not vanish, but rather tilts
at $90^o$ and rotates in the equatorial plane.\\

\begin{figure}
\centerline{
\epsfysize=55mm 
\epsffile{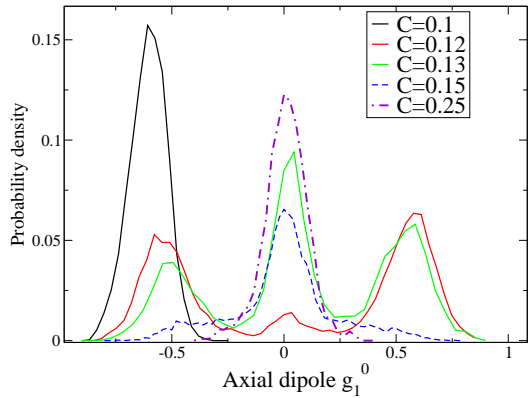}}
\vskip -4mm
\caption{\footnotesize{Probability density function of the axial
    dipolar component at the core-mantle boundary, for different
    values of $C$. Depending on the value of $C$, the distribution can
    be picked around a non-zero value of $g_1^0$ (small $C$, solution
    $D$) or around zero (large $C$, solution $E$). In the bistable
    regime, the distribution can be bimodal or trimodal.}}
\label{PDF}
\end{figure}

In Fig. \ref{time}, the dipolar magnetic field spends
approximatively as much time aligned with the axis of rotation
(solution $D$) as tilted at $90^o$ (solution $E$). In fact, the total
time spent in one state or the other strongly depends on the amplitude
of the symmetry breaking. Fig. \ref{PDF} shows the probability density
function of the dipolar component $g_1^0$ for different values of
$C$. For $C\le 0.1$ (black curve), the equatorial dipole $E$ is not
excited, and only the dipolar configuration $D$ is accessible: the
field does not reverse, and the probability picks around $D$ or $-D$,
depending on the initial conditions. When $C$ is slightly increased,
the system starts to briefly explore the equatorial dipolar state, in
addition to $D$.  The PDF is thus characterized by a non-zero value at
$g_1^0=0$, corresponding to the solution $E$. By symmetry, this
solution is identically connected to $D$ or $-D$, allowing the axial
dipole to reverse the sign of its polarity. For $0.1<C<0.2$, the
probability density function of the axial dipole is then
trimodal. Finally, when $C$ is sufficiently large, only the equatorial
dipole solution $E$ remains, and the probability of $g_1^0$ is
centered around zero.\\

\begin{figure}
\centerline{
\epsfysize=70mm 
\epsffile{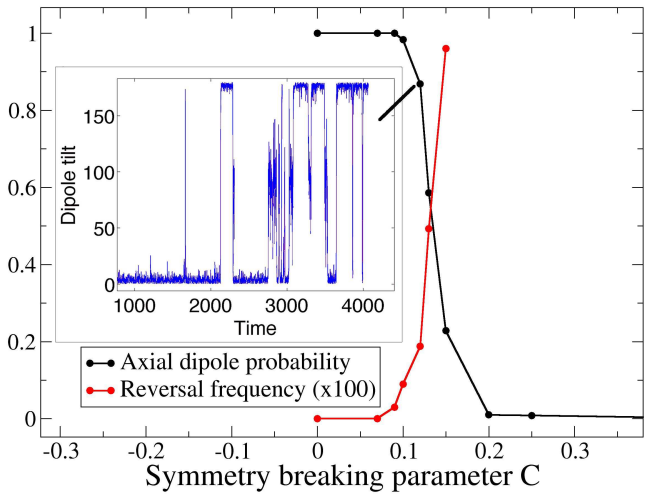}}
\vskip -4mm
\caption{\footnotesize{Black: probability for finding the axial
    dipolar solution, as a function of $C$. The transition from an
    axial to an equatorial dipole field is very sharp. Red: Reversal
    frequency of the axial dipole. As $C$ increases, the basin of
    attraction of the equatorial dipole extends, allowing for more and
    more reversals of the axial dipole. Inset: Time evolution of the
    dipole tilt $\theta_D$ for $C=0.12$: the system spends a very weak
    portion of time in the non-axisymmetric state, and 'Earth-like'
    reversals can be obtained. }}
\label{prob_freq}
\end{figure}

During this transition from a non-reversing dipolar magnetic field to
an oscillating $m=1$ mode, one can also study the direction of the
dipole (Fig. \ref{prob_freq}). The black curve shows the probability
$P_D$ of finding the system in the axial configuration (more
precisely, $P_D$ is defined as the probability that
$\sin(\theta_D)<0.25$, where $\theta_D$ is the dipole tilt angle). The
transition is very sharp, the axial dipole probability dropping
abruptly from one to zero for $C>0.1$. On the contrary, the
probability of finding the equatorial dipole ($\sin(\theta_D)>0.75$)
rapidly increases from zero to one when $C$ is increased. The red
curve shows the reversal frequency of the dipolar solution $D$ versus
the symmetry breaking $C$. When $C$ is increased, the connection with
the attractor $E$ corresponding to the equatorial solution is
larger. Consequently, the connections between the two opposite states
$D$ and $-D$ are more frequent, and the number of reversals increases.

For $C\sim 0.1$, at the very beginning of this transition, the system
spends a long time in the solution $D$. It still explores the
equatorial configuration, but only for a very brief moment during
reversals or excursions. In this case, the distribution tends to be
bimodal (red curve, Fig. \ref{PDF}), despite the fact that three
stable states are involved in the reversal. For instance, the inset of
Fig. \ref{prob_freq} shows the time evolution of the dipole tilt for
$C=0.12$ and illustrates how a weak equatorial symmetry breaking can
produce 'Earth-like' reversals, with a bimodal distribution and a
dipole tilt rapidly switching from $0^o$ to $180^o$.

It is possible to give a naive picture of this mechanism using the
analogy with a heavily damped particle in a tristable potential (a
different but close mechanism is described in \cite{Hoyng01} by
picturing the geodynamo as a bistable oscillator): most of the time,
the system is trapped inside one of the wells (corresponding to $D$ or
$-D$). Due to turbulent fluctuations, the system eventually escapes
one of these stable minima to reaches the opposite one. Between these
two opposite states, there is a third stable potential well, the
equatorial dipole $E$, which creates a connection between $D$ and
$-D$. As $C$ is increased, an exchange of stability takes place from
the potential wells $\pm D$ toward $E$, and reversals become more
frequent (for $C>0.2$, when only $E$ persists, the axial dipole simply
fluctuates around zero). Simply stated, reversals of the axial dipolar
field thus rely on the presence of the equatorial dipole, which is used as
a transitional field during each reversal.\\

Interestingly, this scenario shares strong similarities with the
mechanism for reversals observed in classical geodynamo simulations:
when an homogeneous heat flux is used, reversals of the dipole field
are only observed within a particular transition region of the
parameter space, between a regime in which the field is strongly
dipolar and a regime strongly fluctuating characterized by a
multipolar magnetic structure \cite{Olson06}, \cite{Schrinner10}. In
this case, reversals also result from a bistability between the dipole
and another mode (the multipolar mode), similarly to what happens here
with the equatorial dipole. As in our case, 'Earth-like' reversals are
obtained only if the system is chosen inside the transition region,
but only at the very beginning of this transition, close to the
boundary with the dipolar regime.

Although based on a different mechanism, the behavior of the magnetic
field also has interesting similarities with the model proposed in
\cite{Petrelis09}: reversals are triggered by the equatorial symmetry
breaking, and result from the interaction between the so-called dipole
and quadrupole families of the magnetic field. The intriguing
generation of a strongly hemispherical solution at very small symmetry
breaking is also predicted by this model \cite{Gallet09}. In fact,
depending on the parameters, this model can lead to an hemispherical
solution like the one reported here, or yields polarity reversals
through a saddle-node bifurcation. However, numerical simulations have
shown that this later mechanism is rather selected at sufficiently small
$Pm$ \cite{Gissinger10}, whereas the simulations reported here are
carried at $Pm=20$. Although small $Pm$ simulations are numerically
challenging, it would be interesting to study how the mechanism
described in this letter is modified as $Pm$ is decreased towards more
realistic values.\\

To summarize, we have shown that an equatorial dipole solution can be
generated in geodynamo simulations when the equatorial symmetry of the
flow is broken by an heterogeneous heating at the core-mantle
boundary. Moreover, for weak symmetry breaking, a bistable regime
between this equatorial dipole and the axial dipole is
obtained. Finally, this bistability leads to an interesting scenario
for geomagnetic reversals: The symmetry breaking, by stabilizing the
equatorial dipole, provides the system with a new solution for
connecting the two axial dipole polarities, and sufficiently strong
turbulent fluctuations trigger chaotic reversals of the field. During a
reversal, the transitional field is strongly hemispherical in the bulk
of the flow, and corresponds to an equatorial dipole field at the
core-mantle boundary, rotating around the $z$-axis. In agreement with
paleomagnetic observations, the reversal frequency is directly related
to the equatorial asymmetry of the flow.

\begin{acknowledgments}
We are grateful to Stephan Fauve and Francois Petrelis for their
uncountable comments and useful discussions. This work was supported
by the NSF under grant AST-0607472, the NSF Center for Magnetic
Self-Organization (PHY-0821899) and the ANR Magnet project.
\end{acknowledgments}


\end{document}